\documentclass[twocolumn]{rbef}

\usepackage{amsmath}  
\usepackage{amsfonts} 
\usepackage{amssymb,bm,mathtools}
\usepackage{caption}
\usepackage{subcaption}
\usepackage{siunitx}
\usepackage{calligra}
\DeclareMathAlphabet{\mathcalligra}{T1}{calligra}{m}{n}
\DeclareFontShape{T1}{calligra}{m}{n}{<->s*[2.2]callig15}{}
\newcommand{\scriptr}{\mathcalligra{r}\,}

\DeclareRobustCommand{\uvec}[1]{{%
  \ifcsname uvec#1\endcsname
     \csname uvec#1\endcsname
   \else
    \bm{\hat{\mathbf{#1}}}%
   \fi
}}
\def\diffd{\textup{d}}

\usepackage{cite}
\usepackage[svgnames]{xcolor}

\titulocabecalho{Exploring the magnetic field of Helmholtz and Maxwell coils}
\autorcabecalho{García-Farieta \& Hurtado-M\'arquez}

\numeracao{xxxx}
\volume{xx}
\numero{xx}
\ano{2020xx}
\doi{http://dx.doi.org/xxxx}
\tipodeartigo{Artigos Gerais}

	\author[1]{Jorge Enrique Garc\'ia-Farieta}
	\author[1]{Alejandro Hurtado-M\'arquez}
	\affil[1]{Universidad Distrital Francisco Jos\'e de Caldas, Bogot\'a, Colombia\\Grupo de Investigaci\'on FISINFOR.\thanks{\href{emailto:jegarciaf@correo.udistrital.edu.co}{jegarciaf@correo.udistrital.edu.co}}}

\titulo{Exploring the magnetic field of Helmholtz and Maxwell coils: a computer-based approach exploiting the superposition principle}

\begin{document}
\selectlanguage{english}

\begin{primeirapagina}

\begin{center}
\vspace{-12pt}
\small{Recebido em xxx. Aceito em xxx}
\end{center}

\begin{abstract}
Teaching magnetism is one of the most challenging topics at undergraduate level in programmes with scientific background. A basic course includes the description of the magnetic interaction along with empirical results such as the Biot-Savart law's. However, evaluating the magnetic field due to certain current carrying system at any point in space is not an easy task, especially for points in space where symmetry arguments cannot be applied. In this paper we study the magnetic field produced by both Helmholtz and Maxwell coils at all points in space by using a hybrid methodology that combines the superposition principle and an analytical result. We implement a computational approach, that is based on iterating $n$ times the magnetic field produced by a finite current-carrying wire, to evaluate the magnetic field at any point in space for coils arrangements without using advanced calculus. This methodology helps teachers and students to explore the field due to systems with different levels of complexity, combining analytical and computational skills to visualize and analyse the magnetic field. After our analysis, we show that this is an useful approach to emphasize fundamental concepts and mitigate some of the issues that arise when evaluating the magnetic field for systems proposed in introductory physics textbooks. 
\keywords{teaching magnetic field, superposition principle, computational approach}
\end{abstract}

\end{primeirapagina}
\saythanks

\onecolumn

\section{Introduction}

The electromagnetic theory is one of the fundamental pillars of the scientific training at undergraduate level of physicists and engineers. Within the wide variety of topics considered in the course syllabus, magnetism is one of the most fascinating both in science and engineering due to its applications in so many diverse areas, in industry and in academia as well \cite{Jiles}. However, from an educational point of view, this is also a non-easy topic to teach or to learn \cite{Herrmann, Cahyaningrum}, what has motivated different approaches focused on the effectiveness of the methodologies to avoid possible misconceptions. On the other hand, the continuous development of computational tools has completely revolutionized education, providing a powerful way to visualise and analyse `invisible phenomena' such as the magnetic field, making it a much easier problem mathematically than ever before without requiring experimental equipment \cite{Squire, Rutten, Dori, Aleksandrova, Perkins}.\\

Evaluating the magnetic field, at some point in space, produced by electric current elements is one of the aims in introductory courses. It implies that students must develop extended mathematical skills taking care of empirical relations such as the Biot-Savart law. To help students comprehend this topic better, we study from a computational approach and using the superposition principle, the magnetic field created by two systems: the Helmholtz and the Maxwell coils. The paper is organized as follows: in Section \ref{sec:Bfieldloop} we derive the analytical expressions of the magnetic field produced by both coil configurations in all points of space, then we introduce our approach considering the field created by the superposition of $n$ finite wires as alternative to the analytical solution. In Section \ref{sec:wapproach} we present the computational implementation and the numerical analysis. The results about our methodology and its impact in teaching the magnetic field for the coil configurations considered are discussed in Section \ref{sec:discussion}. Finally, we draw our conclusions in Section \ref{sec:conclusions}.

\section{Magnetic Field Due to Coil Configurations}\label{sec:Bfieldloop}
The Biot-Savart law summarizes in a very clever way the experimental results on the force exerted by an electric current on a nearby magnet. Originally, these results were presented in terms of measurements of the torque on a magnet near a long wire, but were later analyzed in terms of the magnetic field produced by each element of the current \cite{ErlichsonHerman}. The Biot-Savart law describes the magnetic field $d\bm{B}$ at any point in space due to an element $d\bm{l}$ of a current-carrying wire as follows:
\begin{equation}\label{eq:Biot-Savart_Law}
    \diffd\bm{B}(\bm{r})=\frac{\mu_0I}{4\pi}\frac{\diffd\bm{l}\times\uvec{e}_r}{r^2},
\end{equation}
where $I$ is the electric current, $\mu_0$ is the vacuum permeability and $\uvec{e}_r$ is an unit vector along $\bm{r}$ that points from  $\diffd\bm{l}$ to any point in space. Since evaluating the magnetic field at each point in space depends on the distance from the electric source, an analytical expression will highly depend on the geometry of the current-carrying wire and hence on how complicated is the integral of Eq.\eqref{eq:Biot-Savart_Law}. In this context, a common exercise consists of evaluating the magnetic field produced by current carrying circular loop at one point on its axis. This is easily solved, as many textbooks presented, by using \emph{i)} the Biot-Savart law, \emph{ii)} the right-hand rule and \emph{iii)} symmetry arguments. However, evaluating the magnetic field on off-axis points of the ring, turns the problem into advanced and out of the goals of the course because of its difficulty level. In addition, the situation becomes more complex when the configuration involves more than one loop, such as in the case of the Helmholtz and Maxwell coils. The latter is not commonly found in traditional textbooks, while the Helmholtz coils are widely discussed.\\

The arrangement of coils is usually introduced in the textbook section on sources of magnetic field because of its variety of applications (for a detailed description see e.g. \cite{Furth,Byun,Shao,Ghaly,Mihailescu,YuKim,FuZhang}). In fact, they play an important role in industry, medical equipment and material characterization, since they can produce an uniform magnetic field in a small region of space. Specifically the Helmholtz coils are the most simple and widely used system to study the magnetic properties of matter, canceling Earths's magnetic field and in laser cooling and trapping, where a field with a uniform gradient is desired. In what follows, we focus on the mathematical description of the coil configurations and the computational approach.\\

\subsection{The Helmholtz and Maxwell coils}\label{subsec:Bexpres}
The superposition principle states that the field created by different sources simply added together as vectors, thus the total field of a coil arrangement can be expressed as the superposition of the field created by each one of the coils. First let's focus on the magnetic field created by a single loop, a circular wire of radius $R$ located in the $xy$ plane that carries a steady current $I$ as shown in Fig. \ref{fig:coils3D}a. In principle, it is relatively simple to evaluate analytically the magnetic field at any point in space since it is reduced to integrate Eq.~\eqref{eq:Biot-Savart_Law}, however it can be problematic for points in space where it is not possible to use symmetry arguments, even if an appropriate coordinate system is chosen. This is a common example/exercise that appears in many introductory textbooks including some simplifications (see e.g. \cite{Serway,Tipler,Halliday,Griffiths}). In the most general case, it is possible to show that after some manipulations of the Biot-Savart law, the magnetic field created by a circular loop in Cartesian coordinates at any point in space is given by \cite{Griffiths,Loop_NASA,DeTroye,Fano}:
\begin{align}
\tiny
&B_x(\bm{r})=\frac{\mu_0IRz}{4\pi}\int_0^{2\pi}\frac{\cos\varphi^\prime}{\scriptr^3}\mathrm{d}\varphi^\prime = \frac{\mu_0I x z}{2\pi\beta\rho^2}\left[\frac{R^2+r^2}{\alpha^2} E(k^2)-K(k^2)\right]\\
&B_y(\bm{r})=\frac{\mu_0IRz}{4\pi}\int_0^{2\pi}\frac{\sin\varphi^\prime}{\scriptr^3}\mathrm{d}\varphi^\prime = \frac{\mu_0Iyz}{2\pi\beta\rho^2}\left[\frac{R^{2}+r^{2}}{\alpha^2}E(k^2)-K(k^2)\right]=\frac{y}{x}B_x \\
&B_z(\bm{r})=\frac{\mu_0IR}{4\pi}\int_0^{2\pi}\frac{R-x\cos\varphi^\prime-y\sin \varphi^\prime}{\scriptr^3}\mathrm{d}\varphi^\prime = \frac{\mu_0I}{2\pi\beta}\left[\frac{R^2-r^2}{\alpha^2}E(k^2)+K(k^2)\right]\label{eq:gralloop}
\end{align}
where
$\scriptr\equiv\left(r^2+R^2-2xR\cos\varphi^\prime-2yR\sin\varphi^\prime\right)^{1/2}$, $\alpha^2\equiv R^2+r^2-2R\rho$, $\beta^2\equiv R^2+r^2+2R\rho$ and $k^2 \equiv 1-\alpha^2/\beta^2$. Here $K(k^2)$ and $E(k^2)$ represent elliptic integrals of first and second kind respectively. As mentioned before, even if the problem is solved in cylindrical coordinates, elliptical integrals appear, being the azimuthal component the only one that vanishes because of symmetry. The fact that math tools from advanced calculus are needed to completely solve the problem, is perhaps the reason why many introductory textbooks only present a very simplified version of the real system, i.e. asking for the magnetic field on an axial point at certain distance from the loop center instead of all points in space. For this particular case, the field expression is simply given by:
\begin{equation}\label{eq:one-loop}
\bm{B}(z\uvec{k})=\frac{\mu_{0} I R^{2}}{2\left(R^2+z^2\right)^{3/2}}\uvec{k}.
\end{equation}
These expressions are useful to evaluate the magnetic field created by coil configurations. Specifically, the Helmholtz coils consist of two circular coils of radius $R$, each with $N$ turns, that are perpendicular to a common axis as shown in Fig. \ref{fig:coils3D}b. They carry equal steady currents $I$ in the same direction such that their axial fields are added to each other and the coil centers are separated by a distance equal to their radius. Setting the origin of the coordinate system in the middle point of the coils, as shown in Fig. \ref{fig:coils3D}b, instead of the center of one them, the magnetic field on the $z$ axis is given by:
\begin{equation}\label{eq:Helmholtz_coils_yaxis}
\bm{B}(z\uvec{k})=\frac{\mu_{0} NI R^{2}}{2}\left\{
\frac{1}{\left[R^{2}+\left(z+\frac{R}{2}\right)^{2}\right]^{3/2}}+
\frac{1}{\left[R^{2}+\left(z-\frac{R}{2}\right)^{2}\right]^{3/2}}\right\}\uvec{k}.
\end{equation}
An important feature of Helmholtz coils is that the resultant magnetic field in the region between the coils is very uniform. This can be shown easily, since $dB/dz$ and $d^2B/dz^2$ are both zero at the point midway between the coils, thus we may then conclude that the magnetic field in that small region is uniform. A rigorous analysis on the the magnetic field homogeneity and their expression for several coil arrangements, including the Helmholtz ones, can be found in \cite{JWang}.\\

On the other hand, the Maxwell coils consist of an arrangement of three circular and coaxial coils separated by certain ratios as illustrated in Fig. \ref{fig:coils3D}c. Following the original Maxwell's design \cite{Maxwell}, the central coil has a radius $R$ while the two side coils have a radius $\sqrt{\frac {4}{7}}R$, being located on either side of the main coil at distance $\sqrt{\frac{3}{7}}R$ from its plane. As in the Helmholtz coils, in the Maxwell's arrangement all the three currents have the same direction, however in this case the electric current follows the ratio $I\equiv I_{central}=\frac{49}{64}I_{outer}$. Under these parameters, the total magnetic field for points located on the axis of the coils is given by the following expression:
\begin{equation}\label{eq:Maxwell_coils_yaxis}
\small
\bm{B}(z\uvec{k})=\frac{\mu_{0}NIR^2}{2}\left\{
\frac{1}{\left(R^2+z^2\right)^{3/2}}+
\frac{7/16}{\left[\frac47R^2+\left(z+\sqrt{\frac{3}{7}}R\right)^2\right]^{3/2}}+
\frac{7/16}{\left[\frac47R^2+\left(z-\sqrt{\frac{3}{7}}R\right)^2\right]^{3/2}}\right\}\uvec{k}.
\end{equation}

\begin{figure}
     \centering
	 \centering
	 \includegraphics[width=\textwidth]{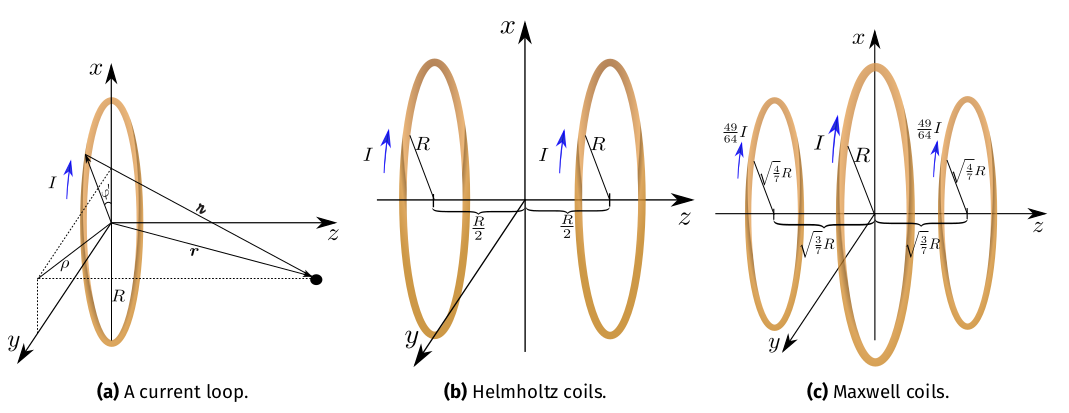}
     \caption{The three different coil arrangements studied in this paper.}
	 \label{fig:coils3D}
\end{figure}

\begin{figure}[h!]
\centering
\includegraphics[scale=0.5]{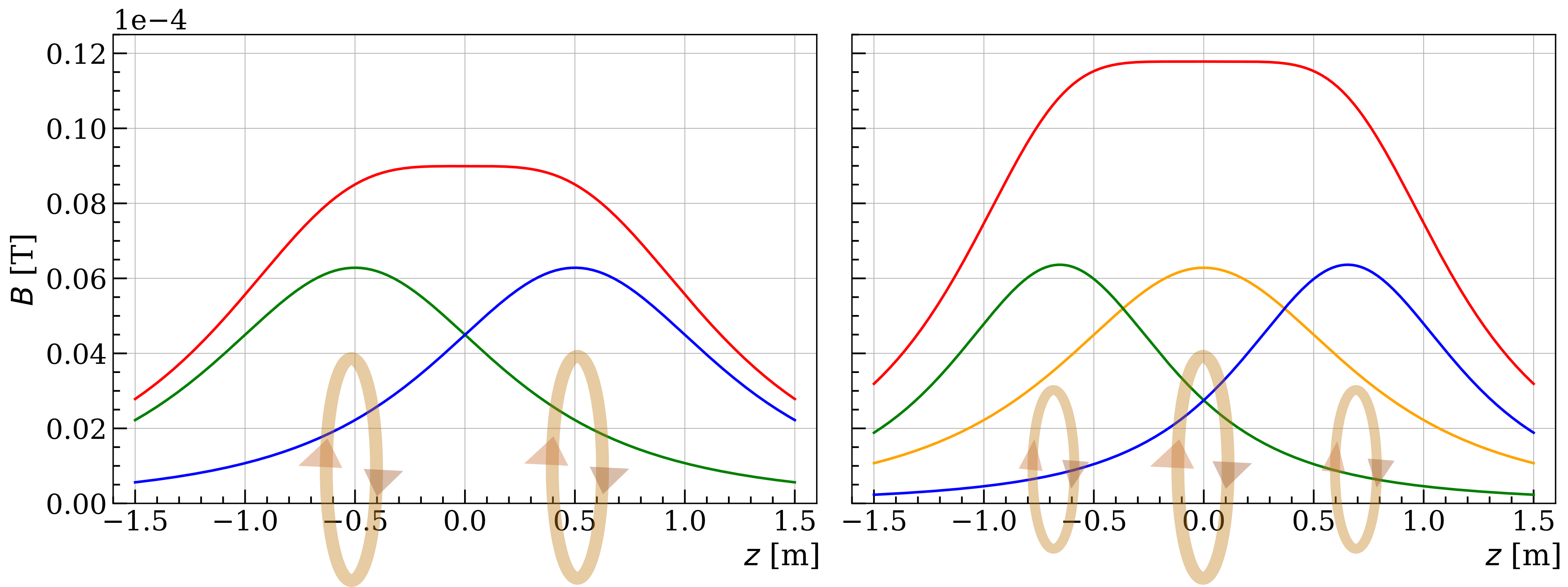}
\caption{The magnetic field evaluated at points along $z$-axis for two coil arrangements: Helmholtz coils (\emph{left panel}) and Maxwell coils (\emph{right panel}). The green, blue and yellow curves show the field produced by the individual coils each system along their coaxial axis. The top red curve is the total magnetic field, illustrating the region of near constant field in the centre.}
\label{fig:CoilsSystems}
\end{figure}

Fig.~\ref{fig:CoilsSystems} shows the magnetic field intensity evaluated at points along $z$-axis created by both the individual coils and the entire system of the Helmholtz and Maxwell arrangements. For illustration purposes we used the following parameters to evaluate Eqs.~\eqref{eq:Helmholtz_coils_yaxis} and \eqref{eq:Maxwell_coils_yaxis} for the Helmholtz coils and keeping the same parameters for the central coil of the Maxwell arrangement: $R=1\,\si{m}$, $N=10$, $I=10\,\si{A}$ and $z\in [-1.5,\,1.5]$ with origin of coordinates in the middle point of the coaxial axis of each system. For these set of parameters the total magnetic field produced by the Maxwell coils is bigger than the produced by the Helmholtz ones, moreover in Fig.~\ref{fig:CoilsSystems} we can appreciate that the system with three coils is more efficient producing an almost uniform distributed magnetic field in the region near the center. These results are only valid for the $z$ axis since Eqs.~\eqref{eq:Helmholtz_coils_yaxis} and \eqref{eq:Maxwell_coils_yaxis} come from simplifying the general expression of a loop by setting $x=0$ and $y=0$ in Eq. \eqref{eq:gralloop}. A more realistic case includes the evaluation of the magnetic field at all points in space, which means that elliptical integrals must be used, however this is not usually commented in textbooks, not even as a warning note or as a suggested advanced exercise. In next section we show how the magnetic field produced by coil arrangements can be evaluated in all points in space using basic mathematical tools that students have within reach, instead of introducing elliptical integrals. This approach is based on the field expression for a finite wire and the superposition principle, moreover it provides an excellent scenario to complement with computational tools.

\subsection{The n-wire approach}\label{subsec:wapproach}

In order to evaluate the magnetic field produced by a loop in all space avoiding elliptical integrals, we approximate the geometry of a loop as a regular $n$-sides polygon. Thus the magnetic field expression is reduced to the superposition of the field produced by $n$ finite wires carrying the same electric current, which is a much simpler problem to solve in the three-dimensional space. The magnetic field due to a finite wire is one of the basic examples introduced in electromagnetism courses, this approximation allows to explore the full solution after applying the Biot-Savart law without loss of generality by exploiting the superposition principle. Moreover, this approach has been explored for several configurations, see e.g. \cite{Garfanwire}, providing a simple evaluation of the field. This has the advantage of being easy to implement numerically since the method iterates a simple geometry to obtain a complex one, keeping the physical properties intact.\\

\begin{figure}[h!]
     \centering
         \includegraphics[width=0.45\textwidth,height=8cm]{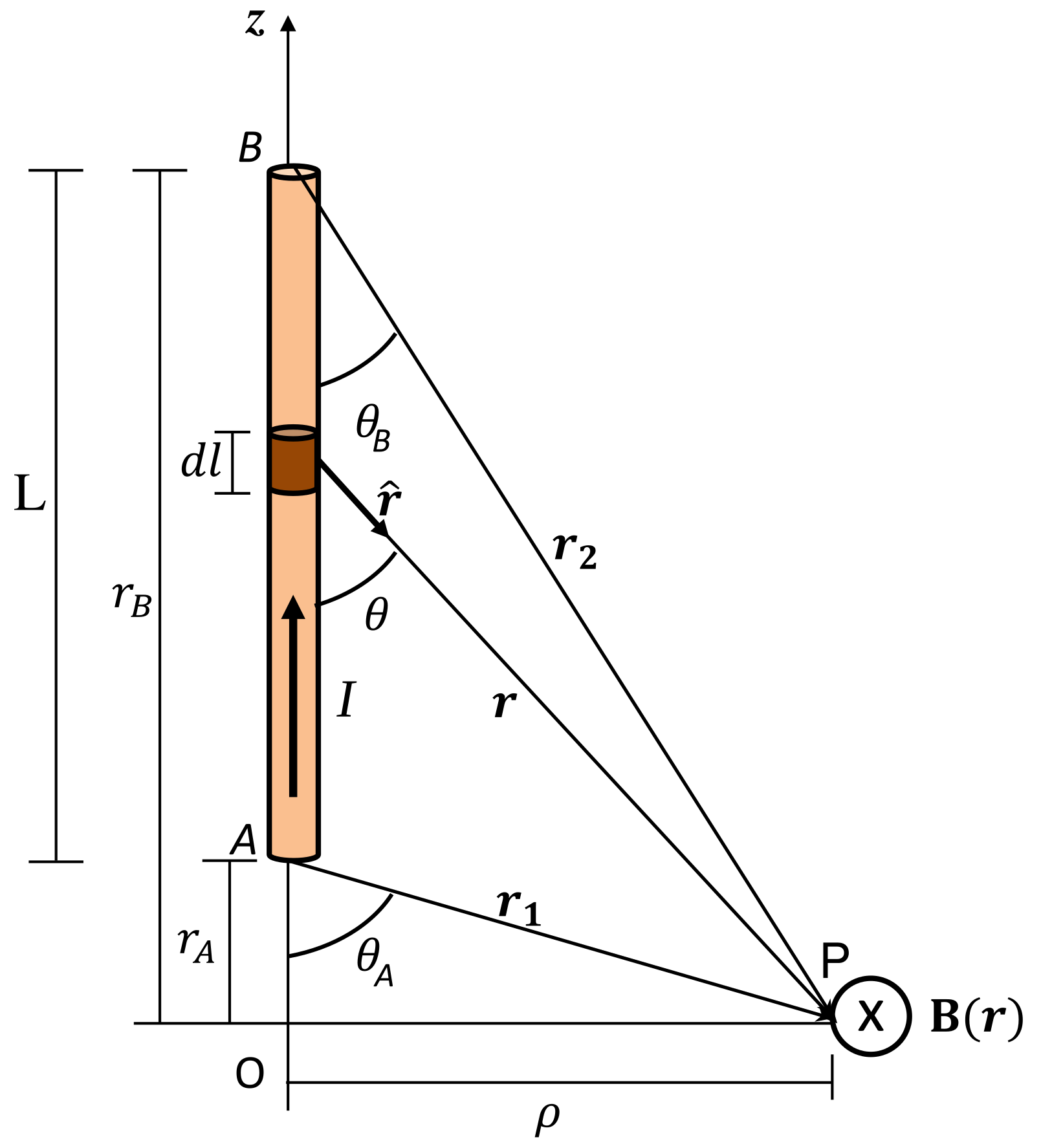}
         \includegraphics[width=0.45\textwidth,height=8cm]{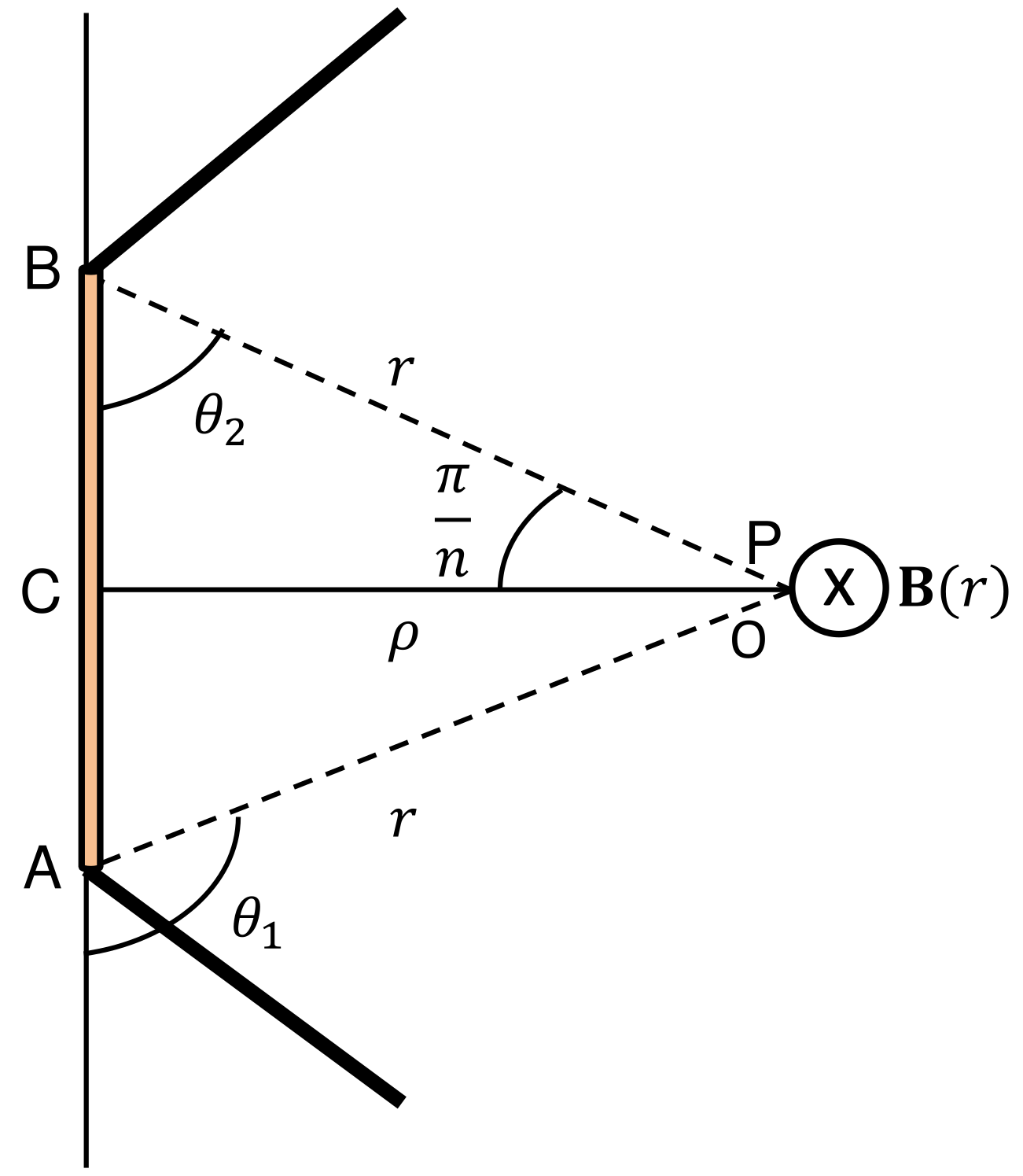}
      \caption{\emph{Left panel:} a finite straight wire carrying a steady current along $z$-axis. The magnetic field is evaluated at the point P. \emph{Right panel:} superposition of $n$ straight wires forming a regular polygonal shaped wire.}
      \label{fig:wire_system_polygon_wire}
\end{figure}

As it is well known, for a thin and straight wire carrying a current $I$ placed along the
$z$-axis, as shown in Fig. \ref{fig:wire_system_polygon_wire}, the magnetic
field at an arbitrary point $P$ is given by:
\begin{equation}\label{eq:Integral_BiotSavart4}
\bm{B}=\frac{\mu_0I}{4\pi\rho}\left(\cos\theta_B-\cos\theta_A\right)
\uvec{e}_\varphi\quad\text{where}\quad\cos\theta_j=\frac{z-z_j}{\sqrt{\rho^2+(z-z_j)^2}}\quad \text{and}\quad j=\{A,B\}.
\end{equation}
where cylindrical coordinates have been used for simplicity to map the
three-dimensional space, taking advantage of the fact that
$d\bm{l}$ and $\bm{r}$ are located in the same plain. Consequently, the
magnetic field is always oriented along $\uvec{e}_\varphi$ for each point $P$, i.e, following concentric circle trajectories regardless of the segment size. The Eq. \eqref{eq:Integral_BiotSavart4} gives a general description of the magnetic field for a current-carrying wire of finite length $L$, however, for the computational implementation it is useful to express it in Cartesian coordinates considering that the wire can be oriented in any direction. Using the law of cosines and vector decomposition, the magnitude of the magnetic field for an arbitrary located wire can be re-written as \cite{YuZhou}:
\begin{equation}\label{eq:Integral_BiotSavart5}
B=\frac{\mu_0I}{4\pi\rho}\left(\frac{r_2^2-r_1^2+L^2}{2Lr_2}-\frac{r_2^2-r_1^2-L^2}{2Lr_1}\right),
\end{equation}
with
$\rho=\sqrt{2r_1^2r_2^2+2r_1^2L^2+2r_2^2L^{2}-r_1^4-r_2^4-L^4}/2L$. On
the other hand, the field direction can be obtained by projecting $\uvec{e}_\varphi$ along the Cartesian
coordinates. From the geometrical construction of Fig.
\ref{fig:wire_system_polygon_wire}(left), and without lost of generality for
an arbitrary located wire, the field orientation can
be written as follows:
\begin{equation}\label{eq:eunit1}
\uvec{e}_\varphi=\frac{\bm{L}\times\bm{r}_1}{\left\|
\bm{L}\times\bm{r}_1\right\|}=\frac{(\bm{r}_B-\bm{r}_A)\times(\bm{r}-\bm{r}_A)}{\left\|
(\bm{r}_B-\bm{r}_A)\times(\bm{r}-\bm{r}_A)\right\|},
\end{equation}
with $(\bm{r}_B-\bm{r}_A)\times(\bm{r}-\bm{r}_A)=\begin{pmatrix}
\uvec{i} & \uvec{j} & \uvec{k}\\
x_B-x_A & y_B-y_A & z_B-z_A\\
x-x_A   & y-y_A   & z-z_A\\
\end{pmatrix}$. This expression can be simplified as:
\begin{equation}\label{eq:eunit2}
\uvec{e}_\varphi=\frac{v_x}{\sqrt{v_x^2+v_y^2+v_z^2}}\uvec{i}+\frac{v_y}{\sqrt{v_x^2+v_y^2+v_z^2}}\uvec{j}+\frac{v_z}{\sqrt{v_x^2+v_y^2+v_z^2}}\uvec{k},
\end{equation}
where $v_x = (y_B-y_A)(z-z_A)-(y-y_A)(z_B-z_A)$, $v_y =
(x-x_A)(z_B-z_A)-(x_B-x_A)(z-z_A)$ and $v_z =
(x_B-x_A)(y-y_A)-(x-x_A)(y_B-y_A)$. Setting $v=\sqrt{v_x^2+v_y^2+v_z^2}$, the Cartesian components of the magnetic field due to a
finite wire, at any point in space are given by:
\begin{equation}\label{eq:Bfield_components}
\begin{gathered}
B_x = B\frac{v_x}{v}\\
B_y = B\frac{v_y}{v}\\
B_z = B\frac{v_z}{v}.
\end{gathered}
\end{equation}

To illustrate the idea behind the $n$-wire approach, let's consider a
$n$-side regular polygon carrying a current $I$. Since each side
corresponds to a finite wire, then Eq. \eqref{eq:Integral_BiotSavart4}
can be used to find $B$ at any point of space. For instance, the field at the center of the polygon can be obtained using Fig. \ref{fig:wire_system_polygon_wire}, where $AB$ is one of the sides and $\rho$ is the radius of the inscribed circle. The angles $\theta_2$ and $\theta_1$ are derived from the geometrical construction, taking into account that the inner angle of the polygon is given by $\angle AOB=2\pi/n$, thus $\angle BOC=\pi/n$. Considering the $n$ wires and using angle properties, the magnitude of the magnetic field in the center of a circular loop of radius $r$ is recovered when $n\rightarrow\infty$ and $\rho=r$, resulting
\begin{eqnarray}\label{eq:Bfield_polygon}
B&=&\frac{\mu_0NI}{4 R}\lim_{n\rightarrow\infty}\frac{n}{\pi}\left[\cos\left(\frac{\pi}{2}-\frac{\pi}{n}\right)-\cos \left(\frac{\pi}{2}+\frac{\pi}{n}\right)\right],\nonumber\\
&=&\frac{\mu_0NI}{2R}\lim_{n\rightarrow\infty}\frac{n}{\pi}\sin\left(\frac{\pi}{n}\right),\nonumber\\
&=&\frac{\mu_0NI}{2 R}.
\end{eqnarray}

\section{Numerical analysis}\label{sec:wapproach}

The numerical pipelines to evaluate the magnetic field due
to an arbitrarily oriented finite wire have been written in \texttt{python3}
language\footnote{The Jupyter notebook to reproduce the figures of this paper is freely available on the author's GitHub repository \url{https://github.com/jegarfa/Magnetic-field-Helmholtz-Maxwell-coils}.}. The implementation consists of defining two functions to \emph{i)} compute the magnitude of the magnetic field using Eq. \eqref{eq:Integral_BiotSavart5}, and \emph{ii)} evaluate the Cartesian components of the field following Eqs. \eqref{eq:eunit2} and \eqref{eq:Bfield_components}. In order to compute the magnetic field produced by a current carrying system at any point in space, it is enough to define its geometry as a superposition of $n$ finite wires and then iterate the routine that evaluates the magnetic field over each wire. In our case, such a system is an arrangement of coils modelled as $n$ side polygons, which satisfy the characteristics of the Helmholtz and Maxwell coils.\\

\begin{figure}[hbt]
\centering
\includegraphics[width=0.6\textwidth]{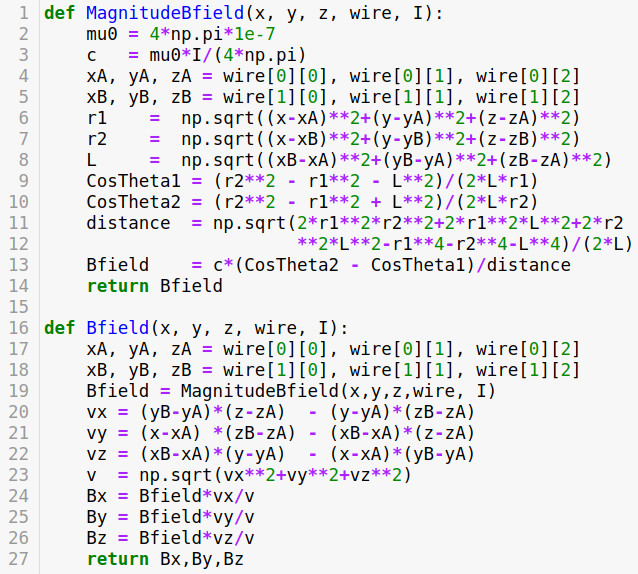}
\caption{Python snippet of the $n$-wire approach to evaluate magnetic fields at any point in space.}
\label{fig:pythoncode}
\end{figure}

\begin{figure}[hbt]
     \centering
         \centering
         \includegraphics[width=0.32\textwidth, height=4cm]{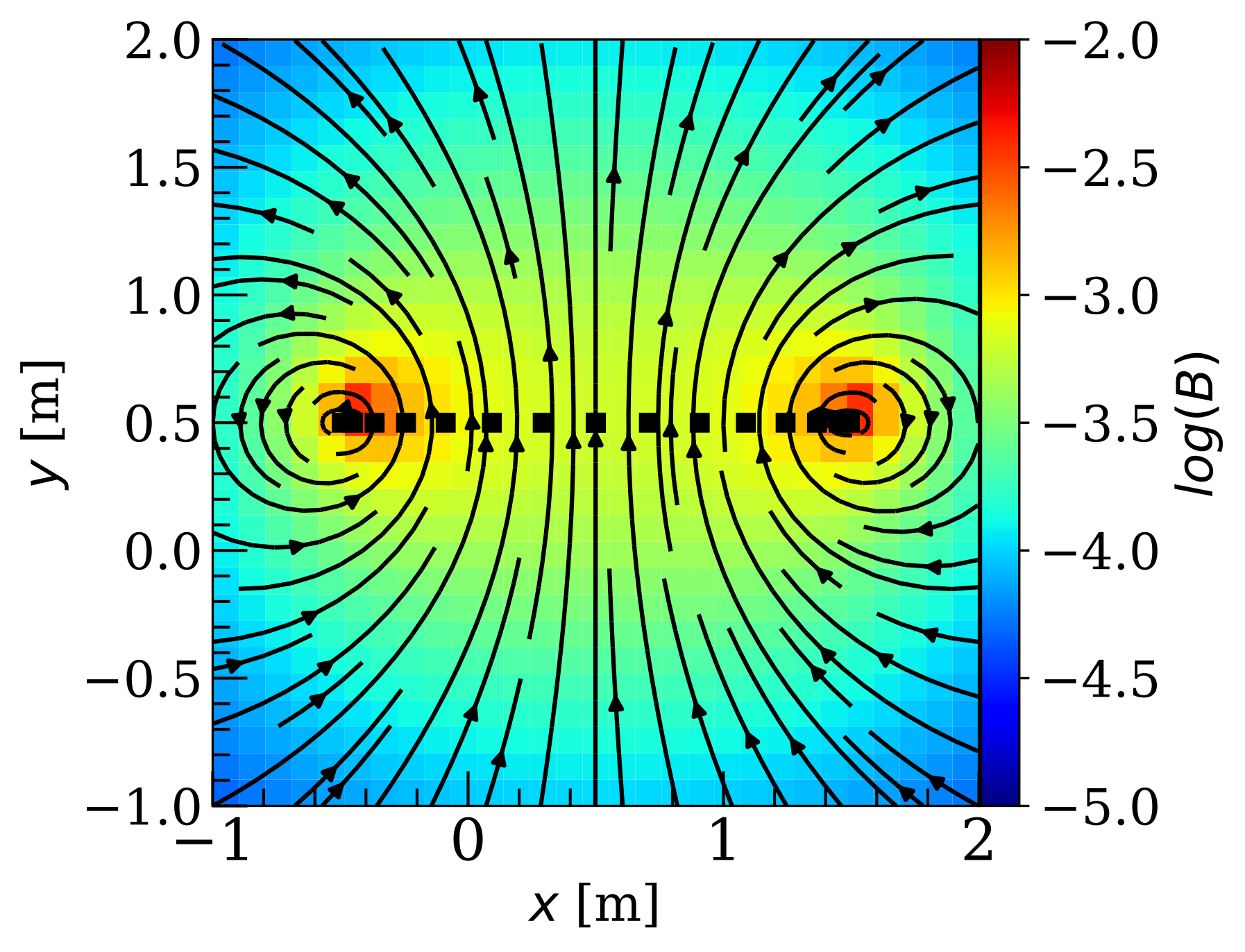}
         \includegraphics[width=0.32\textwidth, height=4cm]{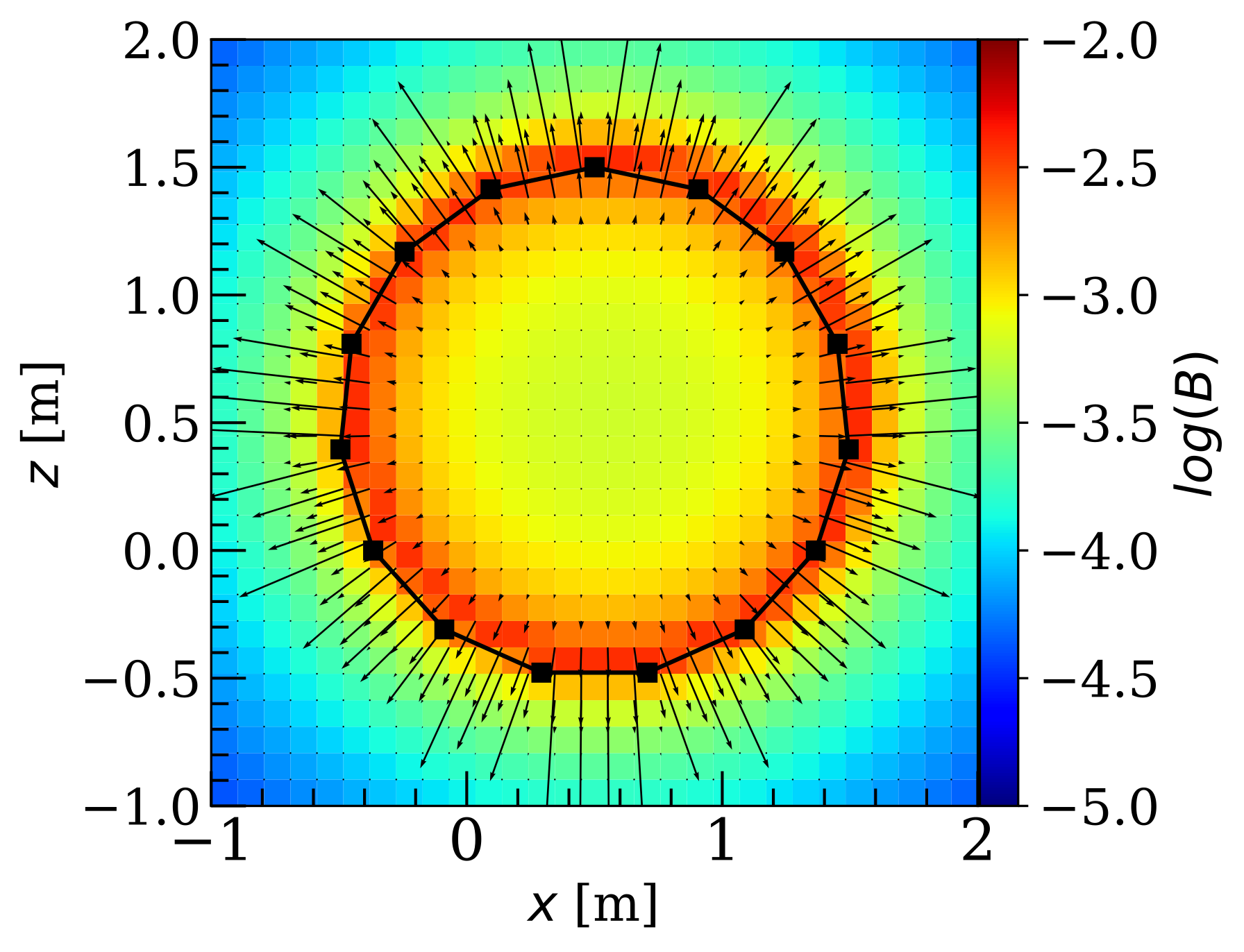}
         \includegraphics[width=0.32\textwidth, height=4cm]{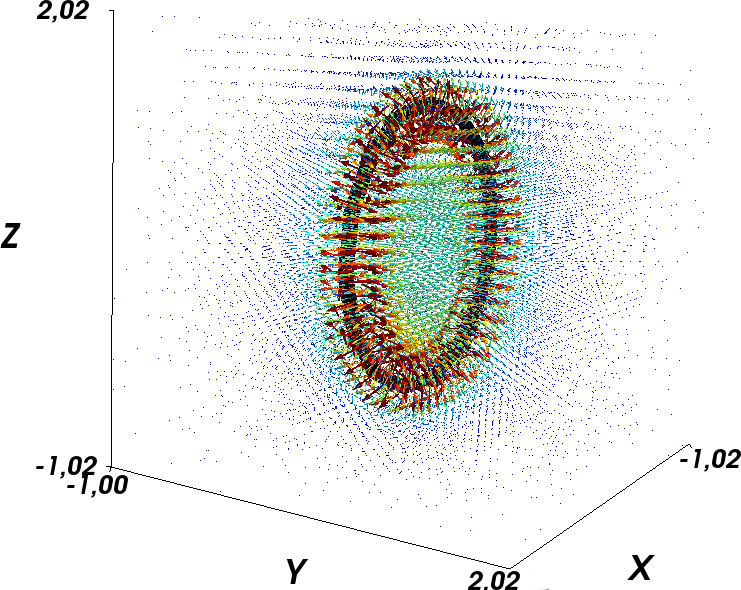}         
      \caption{2D and 3D projections of the magnetic field created by a circular loop evaluated with the $n$-wire approach.}
      \label{fig:coils3D}
\end{figure}

Figure \ref{fig:pythoncode} displays a snippet of the code that implements the $n$-wire approach to evaluate magnetic field due to a system that can be expressed geometrically as a superposition of $n$ wires. Using this approach, we first evaluated the magnetic field produced by a single circular loop in all points in space within a 3D grid. Fig. \ref{fig:coils3D} illustrates the results for a single loop approximated by a regular polygon of $15$ sides inside a meshgrid of $50^3$ points, with radius $r=1~m$ and centred at the point $(x,y,z)=(0.5,0.5,0.5)~m$. The figure shows the 2D projections in the $xy$ and $xz$ planes and the 3D representation of the magnetic field vectors in each point of the grid.\\
\begin{figure}
     \centering
	 \includegraphics[width=\textwidth]{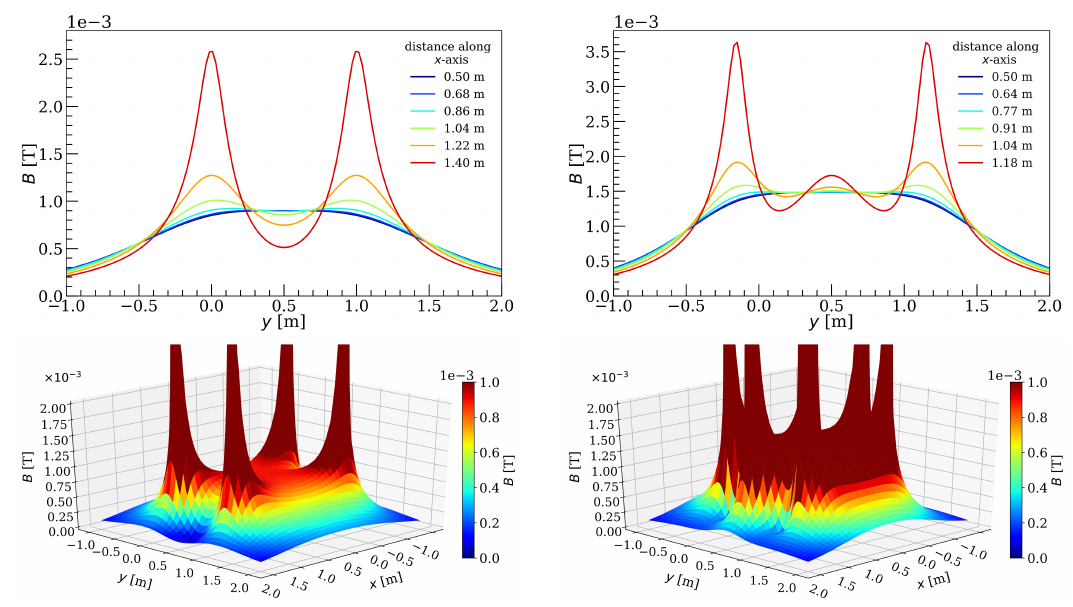}         
     \caption{\emph{Upper panels:} magnetic field intensity along axes parallel to the central one in the plane $z=0$.
     \emph{Lower panels:} magnitude of the magnetic field in the plane $z=0$ as a function of the spatial coordinates. The plots on the left side correspond to the Helmholtz coils while the ones on the right side correspond to the Maxwell coils.}
     \label{fig:Coils_2D_Bfield1}
\end{figure}

Then we repeated the analysis for both arrangements: Helmholtz and Maxwell coils, by approximating the circular loops by regular polygons of $80$ sides inside a meshgrid of $50^3$ points. The first of these systems (Helmholtz coils), consists of two identical coils positioned in parallel to each other in the $xz$ plane with centers aligned with the $y$-axis. The coil centers are $(x_1,y_1,z_1)=(0.5,0.0,0.0)~m$ and $(x_2,y_2,z_2)=(0.5,1.0,0.0)~m$ for the first and second coils respectively. As can be noticed both coils are separated by a distance equal to the radius $1~m$. The second coil arrangement (Maxwell coils), consists of three coils positioned in parallel to each other in the $xz$ plane with centers aligned along the $y$-axis. We keep the radius of the middle coil as $r=1~m$ being centred at $(x_m,y_m,z_m)=(0.5, 0.5, 0.0)~m$. The radius of the outer coils are $r=\sqrt{4/7}$m and they centers are located at $(x,y,z)=(0.5,~r/2\pm\sqrt{3/7}r,~0)~m$ respectively. As can be noticed each coil is apart from each other by $\sqrt{3/7}~m$.
~The Fig. \ref{fig:Coils_2D_Bfield1} shows the behaviour of the magnetic field intensity for both configurations: Helmholtz coils (on the left panels) and Maxwell coils (on the right panels). The magnetic field along axes parallel to the central one in the plane $z=0$ is displayed in the upper panels, while the lower panels show $B(x,y)$ as a function of the spatial coordinates in the meshgrid for the plane $z=0$. In both plots we can appreciate how the magnetic field changes according to the distance from the central axis of the coils. This representation in slices of $B$ at different $x$ values confirms that the field passes from a relatively uniform region at $x=0.5~m$ until it reaches the coils position, where the field is very intense as expected. Moreover, at these distances the Maxwell coils provide an even more uniform magnetic field than the Helmholtz coils with a higher intensity at the region $0 \lesssim y [m] \lesssim 1$.\\ 

\begin{figure}
     \centering
	 \includegraphics[width=\textwidth]{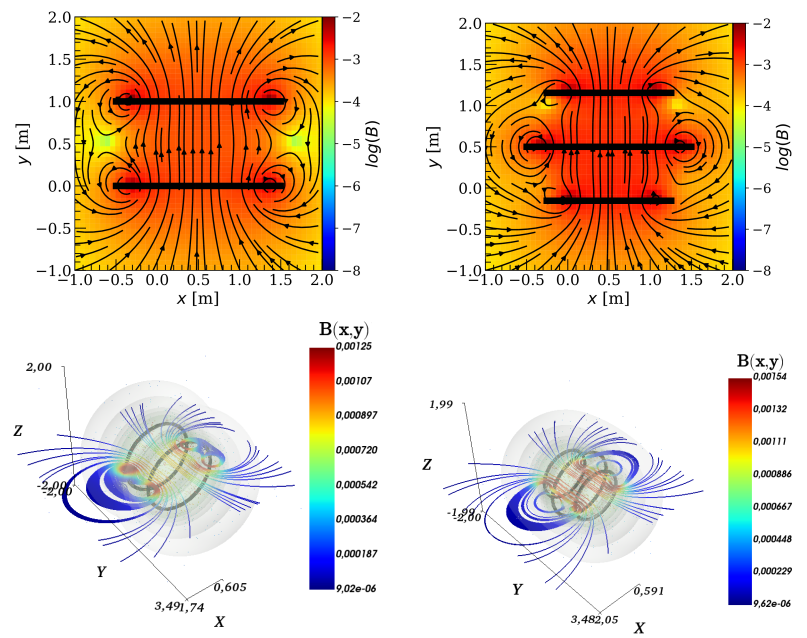}
     \caption{\emph{Upper panels:} magnetic field intensity and 2D patterns of magnetic field lines in the plane $z=0$. \emph{Lower panels:} 3D patterns and contour surfaces of magnetic field lines. The plots on the left side correspond to the Helmholtz coils while the ones on the right side correspond to the Maxwell coils.
     }
     \label{fig:Coils_2D_Bfield2}
\end{figure}

Figure \ref{fig:Coils_2D_Bfield2} shows the 2D iso-contours of the magnetic field lines in the $z=0$ plane and the 3D iso-surfaces of the norm of the field which covers the magnetic field lines produced by the coil arrangements. The plots on the left side correspond to the Helmholtz coils while the ones on the right side correspond to the Maxwell coils. From these patterns we observe that a larger volume with strong homogeneity is achieved with the Maxwell coil as predicted.

\section{Discussion}\label{sec:discussion}

Exploring the magnetic field is a fundamental part of electromagnetism courses where both teacher and students have to face teaching and learning issues. Perhaps, the main issue is related to the fact that evaluating the magnetic field produced by some current carrying systems, at any point in space, require either advanced calculus or an oversimplification of the system. Indeed, many introductory calculus-based electromagnetism textbooks, include symmetry arguments and evaluating the field only at particular points in space (see e.g. \cite{Serway,Tipler,Halliday}). Here appears a first contradiction, since the concept of magnetic field is introduced referring to a quantity that has a value for every point in space but in the exercises it is evaluated only at specific points or regions. Omitting these key points leads students to use incorrect assumptions and algebra tricks when they try to find the magnetic field in more complex systems where same assumptions cannot be applied. An additional challenge to teach magnetism is related to the mathematical tools required to evaluate the field, since it is not always possible to get an analytic expression for any point in space.\\

The superposition principle, combined with an analytical expression of the magnetic field of a finite wire, provides a powerful tool to evaluate the field at any point in space created by several current carrying systems. The approach we implemented in this paper allows to mitigate the above mentioned issues, keeping minimal conceptual assumptions with the advantage of modelling more realistic systems and reducing the gap between the concepts and methods when the field is evaluated. In particular we addressed the results for two very well known coil arrangements, the Helmholtz and Maxwell coils, but the methodology is not limited to them. It can be extended to several configurations as long as they can be expressed in terms of $n$ wires. This hybrid methodology provides a mechanism to explore more complex systems that are close to the real ones involving magnetic field calculations. Moreover, simulations and numerical techniques have the advantage of characterise the physical systems considering a wide range of free parameters, exhibiting properties that otherwise are hidden when there are no analytic solutions. It also encourages students to use computational tools and to create their own systems, which is motivating mainly for engineering students. Finally, the combination of purely mathematical and physical methods (i.e., between the geometrical analysis and the superposition principle in our case study), contributes to the scientific training of future physicists and engineers that must face situations beyond those usually proposed in textbooks. It also gives an additional motivation to students in their learning processes, allowing the electromagnetism course to be influenced by another kind of teaching strategy.

\section{Conclusions}\label{sec:conclusions}

The teaching strategies and resources for physics teaching evolve rapidly with new technologies, specially in topics like magnetism, where physical concepts are not easily assimilable by students. In this paper we have introduced a hybrid method based on superposition principle and computation iterative method to mitigate some of the issues in magnetism teaching. The strategy presented in our work provides a simple method for the calculation of the magnetic fields in several configurations. It also helps understanding the methodology to evaluate the magnetic field due to the Helmholtz and Maxwell coils by using theoretical arguments supported by the Biot-Savart law and numerical techniques.\\

The coil systems studied here help students to understand the Biot-Savart law as a fundamental structure in the calculation of magnetic fields at all points in space, even in those where symmetry arguments cannot be applied. At the same time, this kind of exercises offer a different point of view than those addressed in most of the textbooks used in higher education on the calculation and applications of magnetic fields.\\

In a pedagogical context, designing and implementing computational resources of easy access is perhaps a challenge for educators because it requires more time to integrate the strategy. However, today it is known that these resources become a fundamental tool for the consolidation of scientific and technological literacy processes which can potentially reduce the learning curve of students with scientific background.\\

Extending the methodology proposed here to other topics of electromagnetism can have a positive impact on the way in which fundamental laws of physics are taught such as the Biot-Savart law. In such a case, the traditional teaching techniques based on problem-solving become a learning space where the computational tools are an important factor to achieve meaningful learning from an integrated approach.


\begin{thebibliography}{10}
\bibitem{Jiles} D. Jiles, \textit{Introduction to magnetism and magnetic materials}, (CRC Taylor \& Francis, Iowa, 1998), 2ª ed.
\bibitem{Herrmann} F. Herrmann, Am. J. Phys. \textbf{59}, 5 (1991).
\bibitem{Cahyaningrum} R. Cahyaningrum and A. Hidayat, J. Phys. Conf. Ser. \textbf{1097}, 1 (2018).
\bibitem{Squire} K. Squire, M. Barnett, J.M. Grant and T. Higginbotham, in \textit{Proceedings of the 6th international conference on Learning sciences} (Santa Monica, 2004).
\bibitem{Rutten} N. Rutten, W.R. Van Joolingen and J.T. Van der Veen, Comput Educ. \textbf{58}, 1 (2012).
\bibitem{Dori} Y.J. Dori and J. Belcher, in \textit{Visualization in Science Education}, edited by J.K. Gilbert (Springer, Dordrecht, 2005).
\bibitem{Aleksandrova} A. Aleksandrova and N. Nancheva, Int. J. Inf. Technol. Knowl. \textbf{1}, 1 (2007).
\bibitem{Perkins} K. Perkins, W. Adams, M. Dubson, N. Finkelstein, S. Reid, C. Wieman and R. LeMaster, Phys. Teach \textbf{44}, 1 (2006).
\bibitem{ErlichsonHerman} H. Erlichson, Am. J. Phys. \textbf{66}, 5 (1998).
\bibitem{Furth} H.P. Furth and R.W. Waniek, Nuovo Cimento \textbf{2}, 6 (1955).
\bibitem{Byun} D. Byun, J. Choi, K. Cha, J.O. Park and S. Park, Mechatronics, \textbf{21}, 1 (2011).
\bibitem{Shao} G. Shao and Y.X. Guo, IEEE Trans. Microw. Theory Tech. \textbf{68}, 3 (2020).
\bibitem{Ghaly} S.M. Ghaly and M.O. Khan, Eng. Technol. Appl. Sci. Res. \textbf{9}, 6 (2019).
\bibitem{Mihailescu} B. Mihailescu, I. Plotog and M.N. Velcea, in \textit{2015 IEEE 21st International Symposium for Design and Technology in Electronic Packaging} (Brasov, 2015).
\bibitem{YuKim} C. Yu, J. Kim, H. Choi, J. Choi, S. Jeong, K. Cha, J. Park and S. Park, Sens. Actuator A Phys. \textbf{161}, 297 (2010).
\bibitem{FuZhang} Q. Fu, S. Zhang, S. Guo and J. Guo, Micromachines \textbf{9}, 12 (2018).
\bibitem{Serway} R.A. Serway and J.W. Jewett, \textit{Physics for scientists and engineers with modern physics} (Cengage Learning, Belmont, 2018), 8ª ed.
\bibitem{Tipler} P. Tipler and G. Mosca, \textit{Physics for scientists and engineers} (Macmillan, New York, 2007), 6ª ed.
\bibitem{Halliday} D. Halliday, R. Resnick and J. Walker, \textit{Fundamentals of physics} (John Wiley \& Sons, Jefferson, 2011), 9ª ed.
\bibitem{Griffiths} D.J. Griffiths, \textit{An introduction to electrodynamics} (Pearson, Glenview, 2013), 4a ed.
\bibitem{Loop_NASA} J.C. Simpson, J.E. Lane, C.D. Immer and R.C. Youngquist, \textit{Simple Analytic Expressions for the Magnetic Field of a Circular Current Loop}, available in: https://ntrs.nasa.gov/citations/20010038494.
\bibitem{DeTroye} D.J. DeTroye and R.J. Chase, \textit{The Calculation and Measurement of Helmholtz Coil Fields}, Army Research Laboratory, Adelphi (1994).
\bibitem{Fano} W.G. Fano, R. Alonso and G. Quintana, Elektron \textbf{1}, 2 (2017).
\bibitem{JWang} J. Wang, S. She and S. Zhang, Rev. Sci. Instrum \textbf{73}, 5 (2002).
\bibitem{Maxwell} J.C. Maxwell, \textit{A treatise on electricity and magnetism} (Clarendon Press, Oxford, 1873)
\bibitem{Garfanwire} J.E. Garc\'ia-Farieta and A. Hurtado, \url{arxiv.org/abs/2001.05966} (2019).
\bibitem{YuZhou} Z. Yu, C.H. Xiao, H. Wang and Y.Z. Zhou, Adv. Mater. Res. \textbf{756} (2013).
\end{thebibliography}
\end{document}